\documentclass[fleqn,usenatbib]{mnras}
\usepackage{natbib}
\usepackage[T1]{fontenc}
\usepackage{ae,aecompl}
\usepackage{graphicx}
\usepackage{float}
\usepackage{amssymb}
\usepackage{epsfig}
\usepackage{epstopdf}
\usepackage{subfig}
\usepackage{amsmath}
\usepackage{hyperref}
\usepackage[utf8]{inputenc}
\usepackage{pdflscape}

\title[Radio Dichotomy in High Broadline Quasars]{Radio Dichotomy in Quasars with H$\beta$ FWHM greater than $15,000$ km\,s$^{-1}$  }
\author[Chakraborty et al.]{\parbox{17cm}{Avinanda Chakraborty$^{1,2}$, Anirban Bhattacharjee$^{3}$, Michael S. Brotherton$^{4}$, Ritaban Chatterjee$^{2,1}$, Suchetana Chatterjee$^{2,1}$, Miranda Gilbert$^{3}$}
\vspace{0.2cm}\\
$^{1}${Department of Physics, Presidency University, 86/1 College Street, Kolkata 700073, India}\\
$^{2}${School of Astrophysics, Presidency University, 86/1 College Street, Kolkata 700073, India}\\
$^{3}${Department of Biology, Geology and Physical Sciences, Sul Ross State University, East Highway 90
Alpine, TX 79832, USA}\\
$^{4}${Department of Physics and Astronomy, University of Wyoming,
			Laramie, WY 82071, USA}\\}

\DeclareUnicodeCharacter{2212}{-}

\begin{document}
\maketitle
\begin{abstract}
It has been inferred from large unbiased samples that $10\%$-$15\%$ of all quasars are radio-loud (RL).
Using the quasar catalog from the Sloan Digital Sky Survey, we show that the radio-loud fraction (RLF) for high broad line (HBL) quasars, containing H$\beta$ FWHM greater than $15,000$ km s$^{-1}$, is $\sim 57 \%$. While there is no significant difference between the RL and radio-quiet (RQ) populations in our sample in terms of their black hole mass, Eddington ratio, and covering fraction (CF), optical continuum luminosity of the RL quasars are higher. The similarity in the distribution of their CF indicates that our analysis is unbiased in terms of the viewing angle of the HBL RL and RQ quasars. Hence, we conclude that the accretion disc luminosity of the RL quasars in our HBL sample is higher, which indicates a connection between a brighter disc and a more prominent jet. By comparing them with the non-HBL H$\beta$ broad emission line quasars, we find that the HBL sources have the lowest Eddington ratios in addition to having a very high RLF. That is consistent with the theories of jet formation, in which jets are launched from low Eddington ratio accreting systems. We find that the [O III] narrow emission line is stronger in the RL compared to RQ quasars in our HBL sample, which is consistent with previous findings in the literature, and may be caused by the interaction of the narrow line gas with the jet.

\end{abstract}



\section{Introduction}

Since the very first discovery of quasars \citep{schmidt}, the radio bi-modality of the quasar population has been discussed in the literature \citep[e.g.,][]{white00, brotherton, Mark, cir03, cira03, doi13, tad16, hard20}. While the principal difference between the radio-loud (RL) and radio-quiet (RQ) quasars lies in the strength of their radio jets \citep[e.g.,][]{bridle, mullin}, several studies have been undertaken to address the source of this dichotomy \citep[e.g.,][]{kell, mill, white00, Mark, ive, white07, Behar, zfir, bal12, lixia}. 
The dichotomy has been studied in the context of  black hole (BH) mass \citep [e.g.,][]{Ross}, accretion rate \citep[e.g.,][]{sikora, ham}, host galaxy properties \citep[e.g.,][]{sikora, lag, kim}, as well as black hole (BH) spin \citep[e.g.,][]{blandford, Bford, gar}. It has been proposed that there may be multiple mechanisms responsible for powering the radio emission, for example it may come from a combination of small-scale jets as well as star formation regions \citep{cir03, sab19}.

In the cosmological context, the role of galaxy mergers and host dark matter halos in the RL-RQ dichotomy has also been discussed \citep[e.g.,][]{scott, SY, hick, lag, gar, ham}. Using host galaxy photometry and optical spectroscopy of the nuclear region, several recent works revealed a consistent mass excess of the central BHs powering RL quasars compared to RQ quasars \citep[e.g.,][]{dunlop, Chi, jar}. Other studies have reached similar conclusions \citep[e.g.,][]{mag, Laor, McL} regarding the mass distributions of RL and RQ quasars. 

Some authors have shown that BH mass is correlated with radio luminosity \citep[e.g.,][]{fran, McLu, Ari, Mark, Ross} while some studies report the contrary \citep[e.g.,][]{Ho, oshlack, Woo, Snel}. In the context of this dichotomy, there have been proposals of radio-intermediate sources, which essentially implies the distribution of radio loudness to be quasi-continuous \citep{fal, brotherton, Mark, gur19}. One of the key reasons for which it has been particularly difficult to establish the bimodality, is the absence of a universally accepted definition of radio loudness. A ratio of the radio to optical luminosity has been used most often while the absolute value of the radio luminosity has been assumed to be the characteristic parameter in some other cases. However, for both of those parameters the RL-RQ boundary has not been universal over various studies in the literature \citep[e.g.,][]{bal12}.

\begin{figure}
\begin{center}
 \resizebox{9cm}{!}{\includegraphics{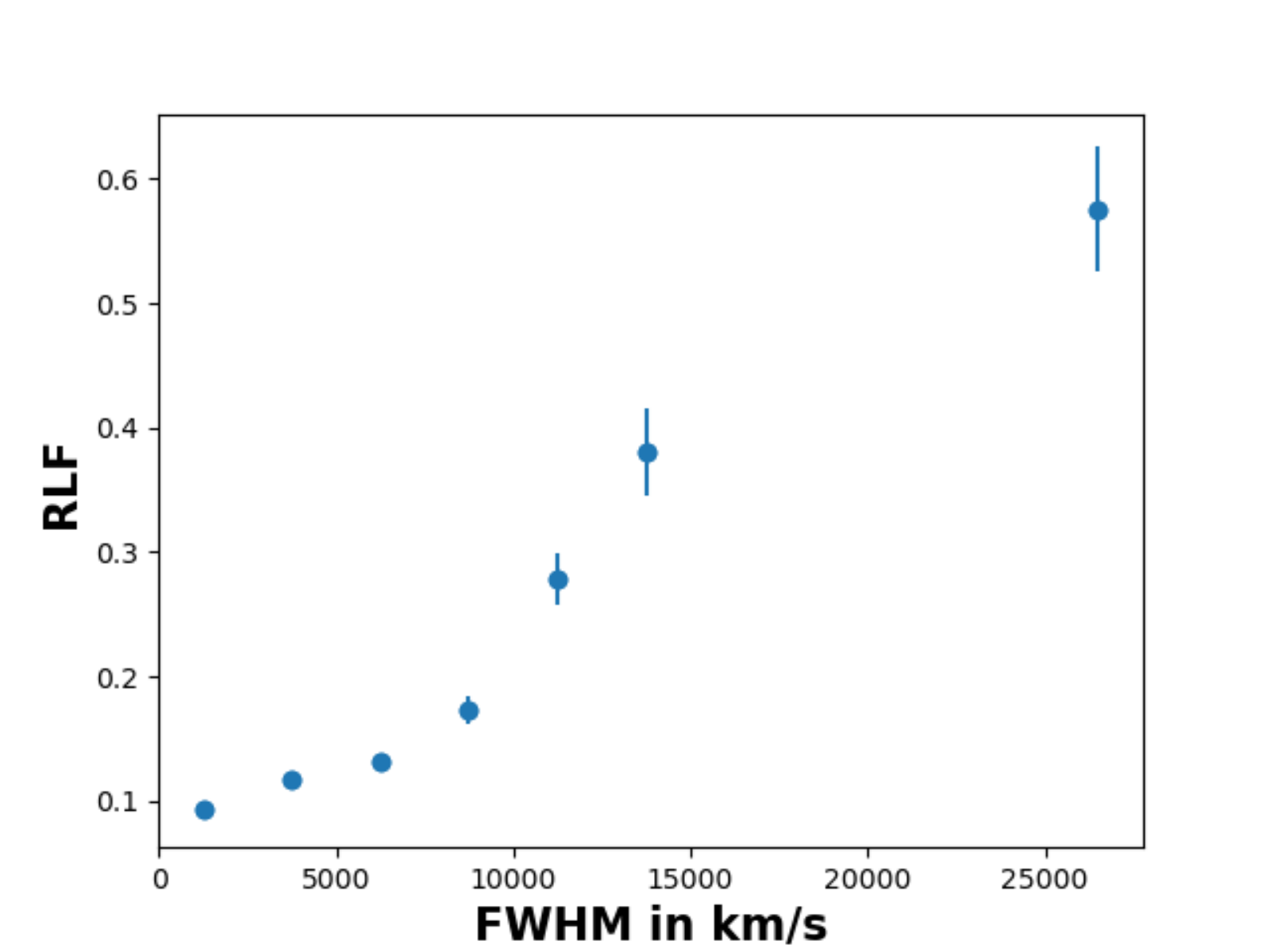}}
\caption{Blue dots denote the radio loud fraction (RLF) as a function of FWHM of broad H$\beta$ emission line. The x-axis represents the central value of the FWHM bin. The error-bars used here are binomial errors of the mean value of the RLF. We see that the RLF increases with FWHM.}
\label{fig:RLF}
\end{center}
\end{figure}

Radio emission in quasars is shown to be related with the optical emission lines \citep[e.g.,][]{Oster}. In radio galaxies, a correlation between emission line luminosities and radio jet luminosities has been established \citep[e.g.,][]{Baum, Saun, raw&sau91, will99, Zir, Butt, Sik}. Therefore, emission line diagnostic has been proposed by some authors as a distinct method to define RL quasars.
However, studies reveal that there can be significant scatter in the radiative vs kinetic emission correlations \citep [see e.g.,][]{raw&sau91, pun&zhang11, mingo14, mingo16, gurkan15}. It has been shown that jets do produce emissions beyond the radio, and are capable of photoionizing the surrounding interstellar medium, but through shock-ionization  they can also produce emission lines \citep[e.g.,][]{min11}.

\cite{zfir} showed that RL quasars are located at a specific region on an H$\beta$ full width half maximum (FWHM) \textit{versus} $R_{Fe}$ plane. We followed \cite{zfir} to define $R_{Fe} = F_{Fe}/F_{H\beta}$, where $F_{H\beta}$ is the optical continuum flux at 5100 \AA  and $F_{Fe}$ is the FeII flux at 4861 \AA~, calculated using the optical continuum flux at 5100 \AA~ and H$\beta$ slope. Among the sample they used roughly 78\% of RL quasars have H$\beta$ FWHM $>$ 4000 km\,s$^{-1}$. While some RQ quasars are located in that part of the plane, the restricted location of RL quasars is statistically significant. 

Accreting BH systems have been found to reside preferentially on a plane in a space comprising disc luminosity, jet power and BH mass \citep[e.g.,][]{Mac, Mer, Tera}. This is known as the fundamental plane of BH activity. It shows that the disc activity, jet emission, and BH mass are connected with each other. However, the exact nature of the above connection is not well understood. In general relativistic magnetohydrodynamic simulations of jet launching and collimation \citep [e.g.,][]{Kom, McK} it is revealed that the conversion of accreted material into collimated jets may depend on multiple variables or a combination thereof, including, BH mass  \citep[e.g.,][] {dunlop, marconi}, BH spin \citep[e.g.,][]{blandford, fernandes}, and disc activity \citep[e.g.,][]{lin, fernandes}. During the accretion process, gas in the disc gets heated and strong ultraviolet (UV) to optical continuum emission is radiated \citep{prin81, malkan82, ma&wang13}. Optical spectroscopic observations of the continuum and line emission of RL and RQ quasars have shed light on the connection between the accretion disc and jet of these systems \citep{kawa}. It has been shown that the radio loud fraction (RLF, hereafter) of quasars decreases with decreasing optical luminosity \citep{jia}.  

In the preliminary work of \citet{chak}, we used a sample of RL and RQ quasars selected through the Sloan Digital Sky Survey (SDSS) and performed a systematic study of their optical continuum and line emission to search for clues in understanding the dichotomy in their radio emission. In particular, we studied the BH properties of RL and RQ objects selected through their H$\beta$~ FWHM. Our results showed that the RLF increases with FWHM, and hence to further understand this result we select a sub-sample of extreme objects in the catalog with H$\beta$ FWHM greater than 15,000 km s$^{-1}$ and study the bolometric luminosity, optical continuum luminosity and fiducial virial BH mass \citep{vester&peter06} distributions of these extreme objects. Gas in the broad line region (BLR) of quasars is photoionized by the emission from the central continuum source at ultraviolet frequencies, and gives rise to prominent recombination lines observed in the quasar spectra \citep{sulen00}. Due to their proximity to the central SMBH the BLR gas has high speed ($10^3 - 10^4$ km\,s$^{-1}$) which results in the Doppler broadening of the emission lines. Thus our sample consists of quasars with large values of the speed of the BLR gas.
\begin{figure}
\begin{center}
\begin{tabular}{c}
        \resizebox{9cm}{!}{\includegraphics{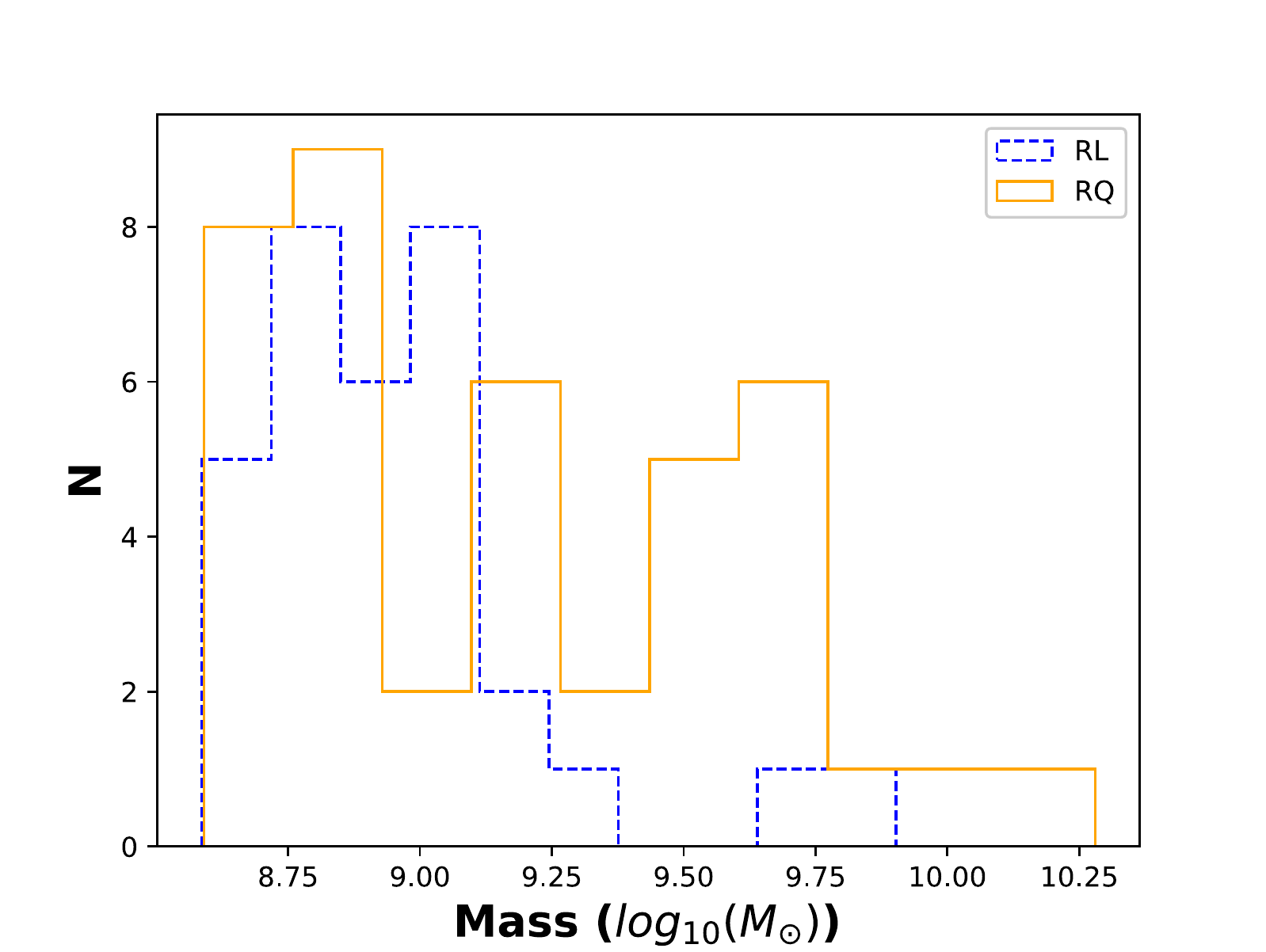}}\\
        \resizebox{9cm}{!}{\includegraphics{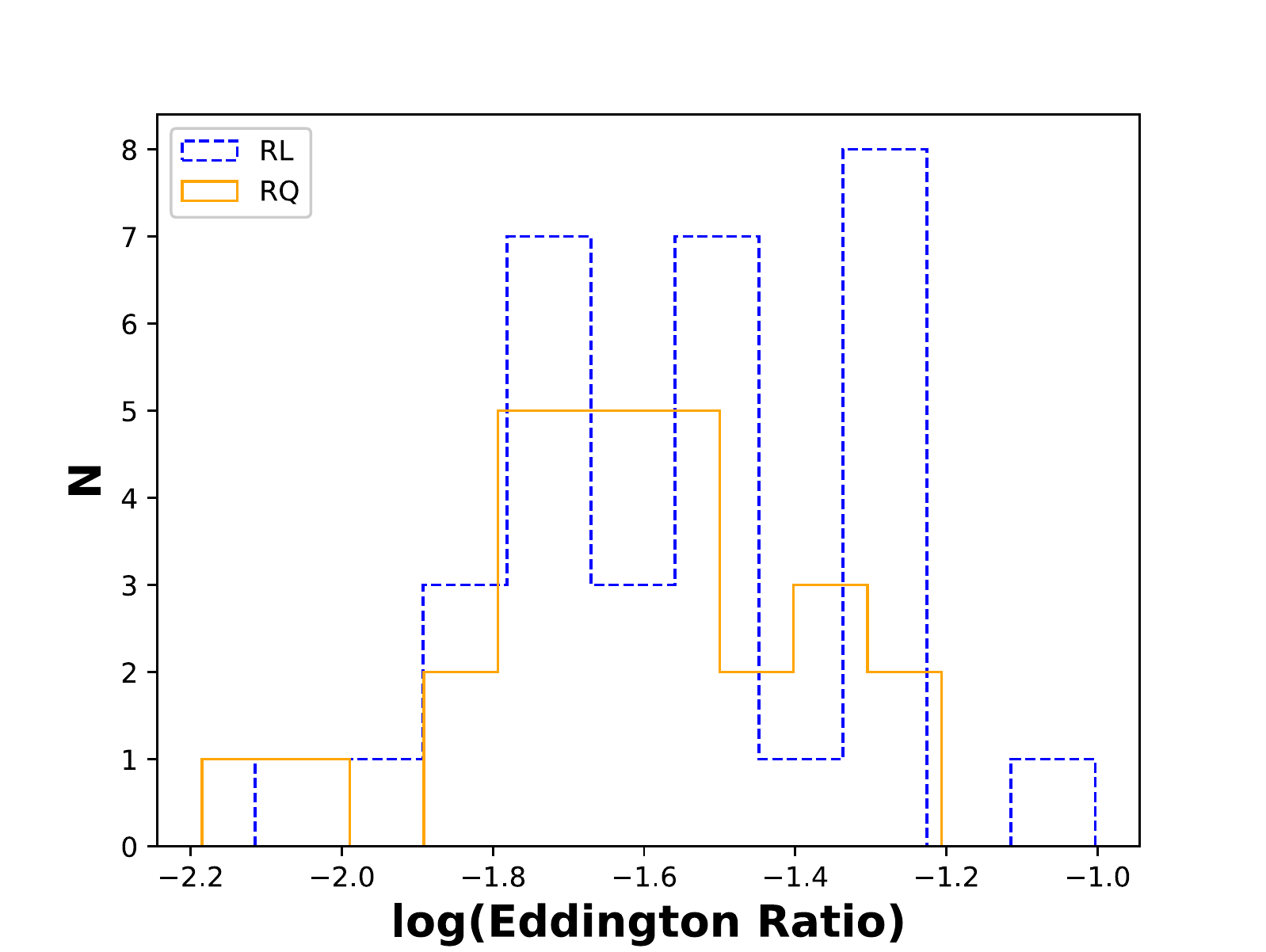}}\\
         \resizebox{9cm}{!}{\includegraphics{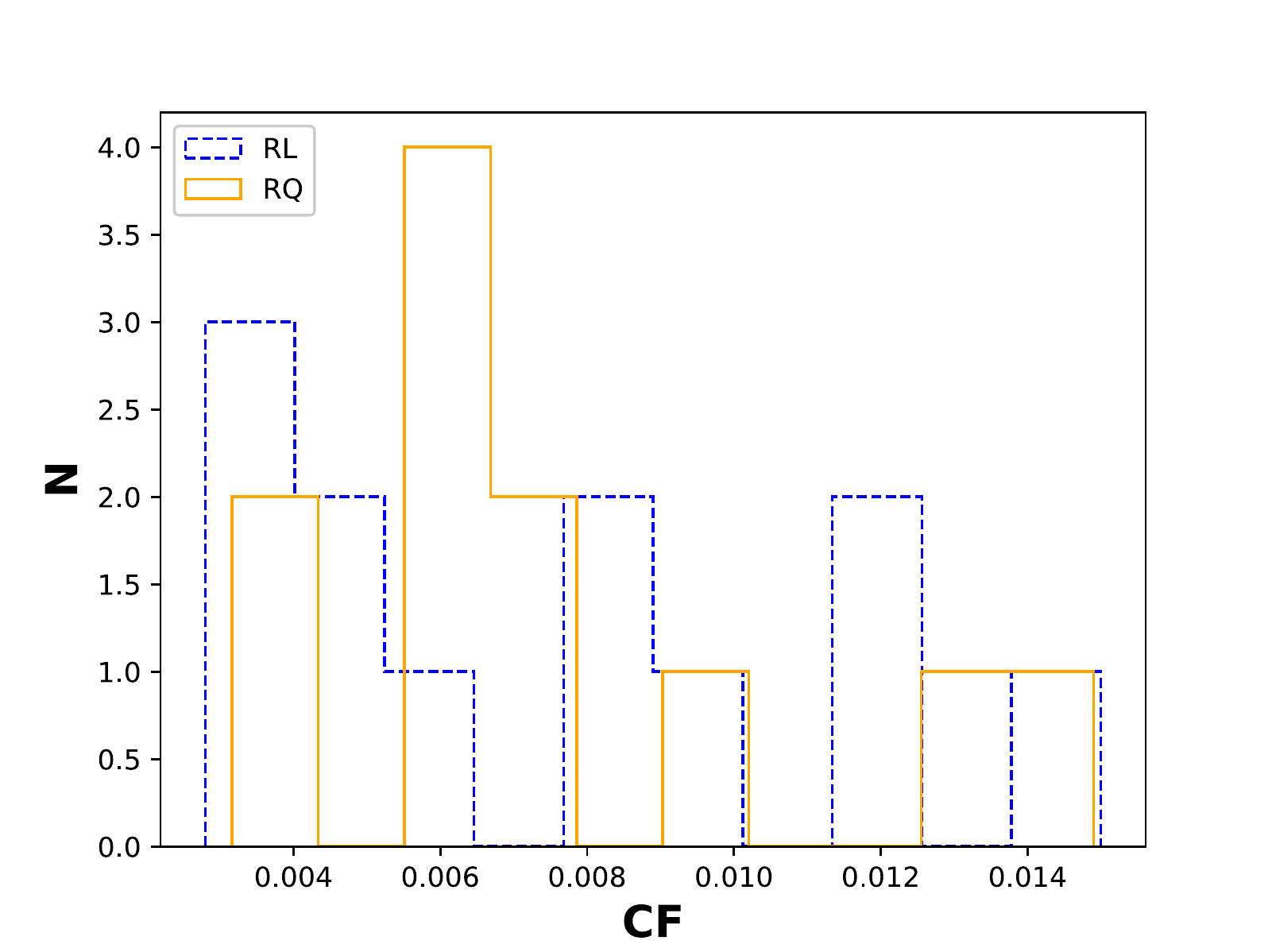}}\\
\end{tabular}
\caption{{\bf Upper Panel: }Orange solid and blue dashed lines denote the distributions of black hole mass of the radio-quiet and radio-loud quasars, respectively, with H$\beta$ FWHM greater than 15,000 km $s^{-1}$ in our sample. {\bf Middle Panel: } Same as the top panel but now for Eddington ratios. {\bf Bottom Panel: } Same as the other panels but for covering fraction. From the figure we note that the distributions of black hole mass, Eddington ratio and covering fraction of the two population are similar.}
\label{fig:M-Ledd-CF}
\end{center}
\end{figure}

In the current work, we investigate whether the radio luminosity, bolometric luminosity, BH mass, and radio loudness parameter of the RL quasars in our sample of high broad line (HBL hereafter; with FWHM greater than $15,000$ km\,s$^{-1}$) quasars are related to each other while comparing them with their relevant RQ counterparts. We construct the composite SDSS spectra of the quasars in our sample and search for any difference between the RL and RQ populations in terms of their mean spectrum. Finally, we compare our HBL sample with the rest of the H$\beta$ broad emission line objects (non-HBL hereafter) in terms of BH mass, Eddington ratio, bolometric luminosity, H$\beta$ FWHM, and $R_{Fe}$ to search for possible clues of higher RLF. 

Throughout this paper we use $\Omega_{\Lambda}$ = 0.7, $\Omega_{0}$ = 0.3 and h = 0.7 as the cosmological parameters \citep{sper}. In \S 2, we give a brief description of our sample and present our results in \S 3. In \S 4 we discuss the results and summarize our conclusions.

\section{Data Sets} 
In this section we discuss our data sets and sample selection. 

\subsection{Sloan Digital Sky Survey (SDSS)}
We use RL and RQ quasars from \citet{shen}, which is the complete quasar catalog obtained from SDSS DR7 \citep{schn}. The catalog consists of $105,783$ spectroscopically confirmed quasars, that are brighter than $Mi = -22.0$ obtained from a sky coverage of $\sim 9,380  {deg}^2$. The catalog consists of quasars that have reliable redshifts and have at least one emission line with FWHM greater than $1,000$ km\,s$^{-1}$. The flux limit for the main spectroscopic sample is $i < 19.1$. Therefore, the majority of quasars are brighter than $i \approx 19$. We have chosen to use DR7 because it includes line fits with multiple Gaussian functions for the broad H$\beta$ lines along with compilation of equivalent widths, line luminosities, and iron emission strengths with high accuracy.

We note that we have checked the SDSS DR10 \citep{paris13} and DR14 \citep{paris18} quasar samples. SDSS DR10 samples do  not have H$\beta$ emission-line properties and SDSS DR14 has very few objects which satisfy our magnitude ($i < 19.1$), redshift ($z$ $<$ 0.75) and FWHM ($>$ 15000 km$s^{-1}$) cuts. Since we are looking for H$\beta$ emission line objects, this redshift limit is necessary. Thus DR7 allows us to construct a larger HBL catalog when compared with DR10 and DR14, making the sample ideal for our analysis. 

\begin{figure*}
\begin{center}
\begin{tabular}{c}
        \resizebox{9cm}{!}{\includegraphics{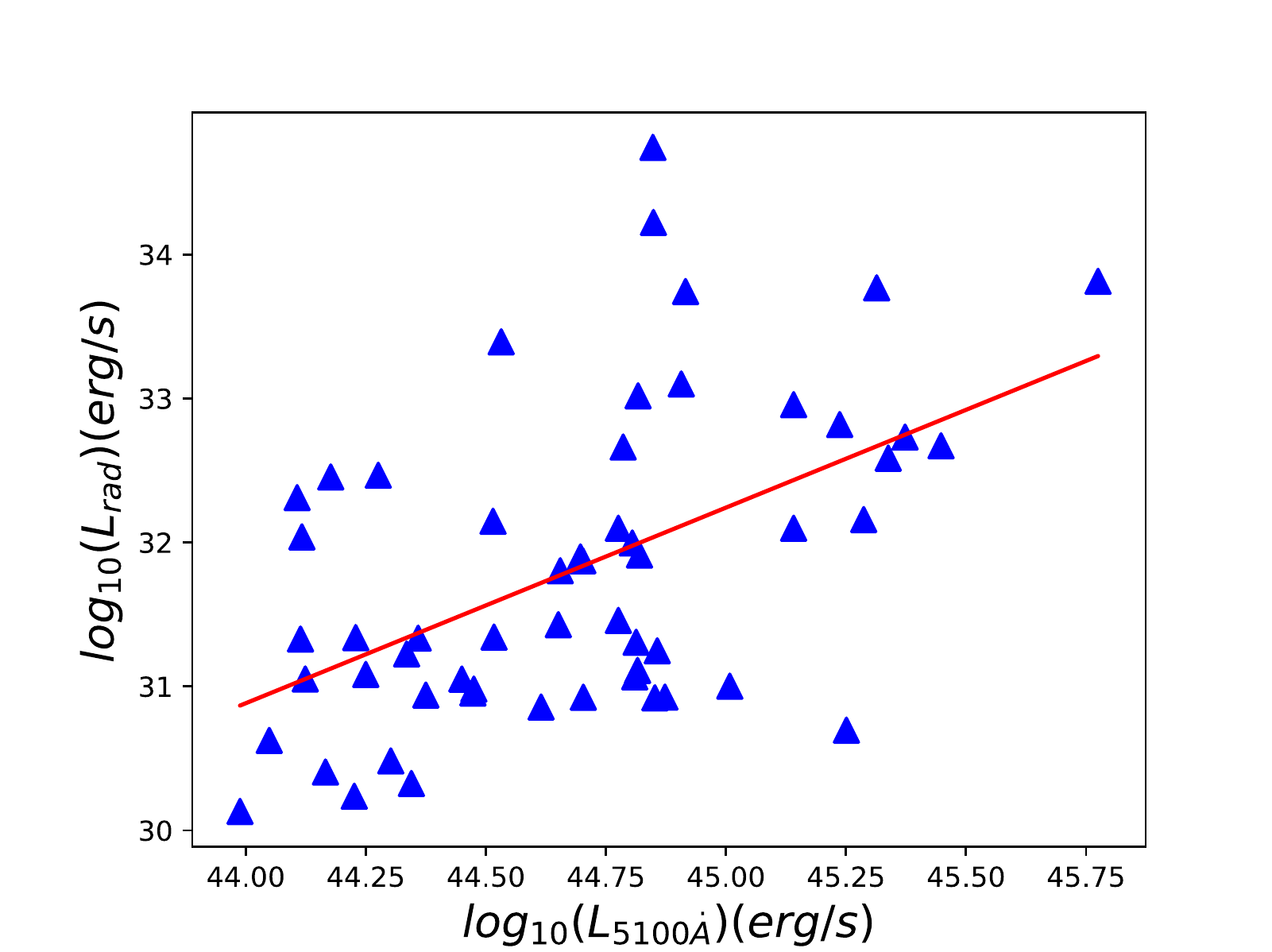}}
        \resizebox{9cm}{!}{\includegraphics{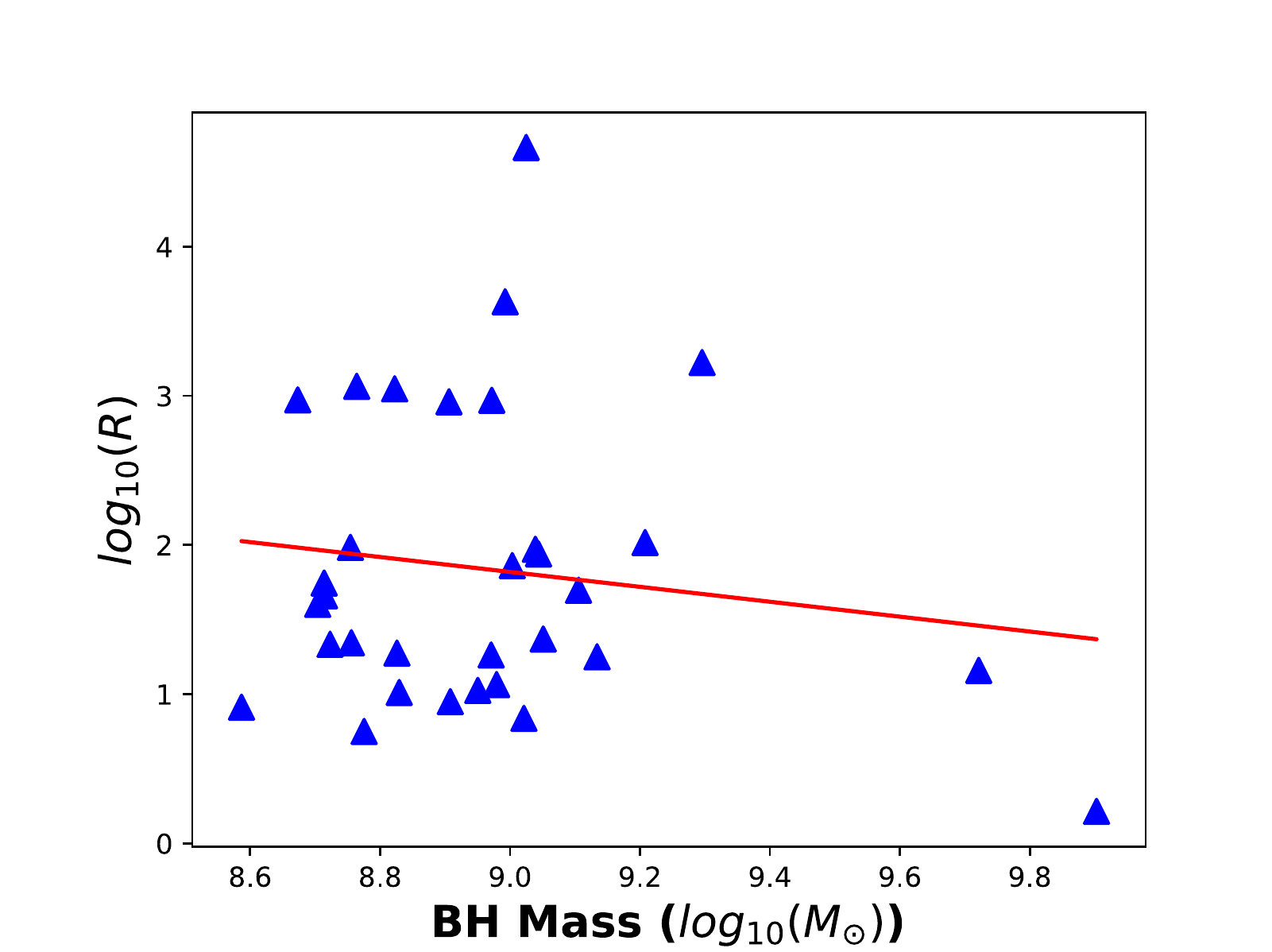}}
\end{tabular}
\caption{{\bf Left Panel: }The blue solid triangles represent the RL quasars in our HBL sample on the radio luminosity \textit{versus} optical continuum luminosity plane. {\bf Right Panel: }The blue solid triangles represent the same sources on the radio-loudness \textit{versus} black hole mass plane.} 
\label{fig:Lrad-L5100-R-M}
\end{center}
\end{figure*}

\subsection{Faint Images of the Radio Sky at Twenty-cm (FIRST)}
Faint Images of the Radio Sky at Twenty-Centimetres (FIRST) is a radio survey \citep{white07} that scanned over 10,000 square degrees of the North and South Galactic Caps at 1.4 GHz. It produces images with $1.8''$ pixels with an rms of 0.15 mJy\,bm$^{-1}$, resolution $5''$ and a threshold flux density of 1.0 mJy\,bm$^{-1}$. FIRST was designed to overlap with the SDSS survey with 40\% identification rate for optical counterparts at SDSS limiting magnitude of $\sim23$.


\cite{shen} use a specific criterion to cross-match the DR7 quasars with the FIRST survey. To include radio properties, following \cite{jia} $R = f_{6cm}/f_{2500}$ has been estimated, where $f_{6cm}$ and $f_{2500}$ are the rest-frame flux densities at 6 cm and 2500\AA, respectively. SDSS sources which have single counterparts within $5''$ (core dominated), or have multiple counterparts within $30''$ (lobe dominated) in FIRST are classified as RL quasars \citep{shen}. Out of all 105,783 quasars in SDSS, 99,182 have a FIRST match. Among those, 9,399 are identified as RL and 88,979 as RQ quasars. We have taken a redshift ($z$ $<$ 0.75) limited sample to consider only the quasars which have broad H$\beta$ emission lines in their spectra \citep{rich}. Our final sample contains 298 RL and 1,910 RQ quasars. Among RL quasars, 56 have HBL (FWHM $>$ 15000 km$s^{-1}$) while the corresponding number in RQ quasars is 41. We note that the redshift distributions of RL and RQ quasars in our sample are similar \citep [CB21 hereafter]{chak}.

\begin{figure*}
\begin{center}
\begin{tabular}{c}
        \resizebox{9cm}{!}{\includegraphics{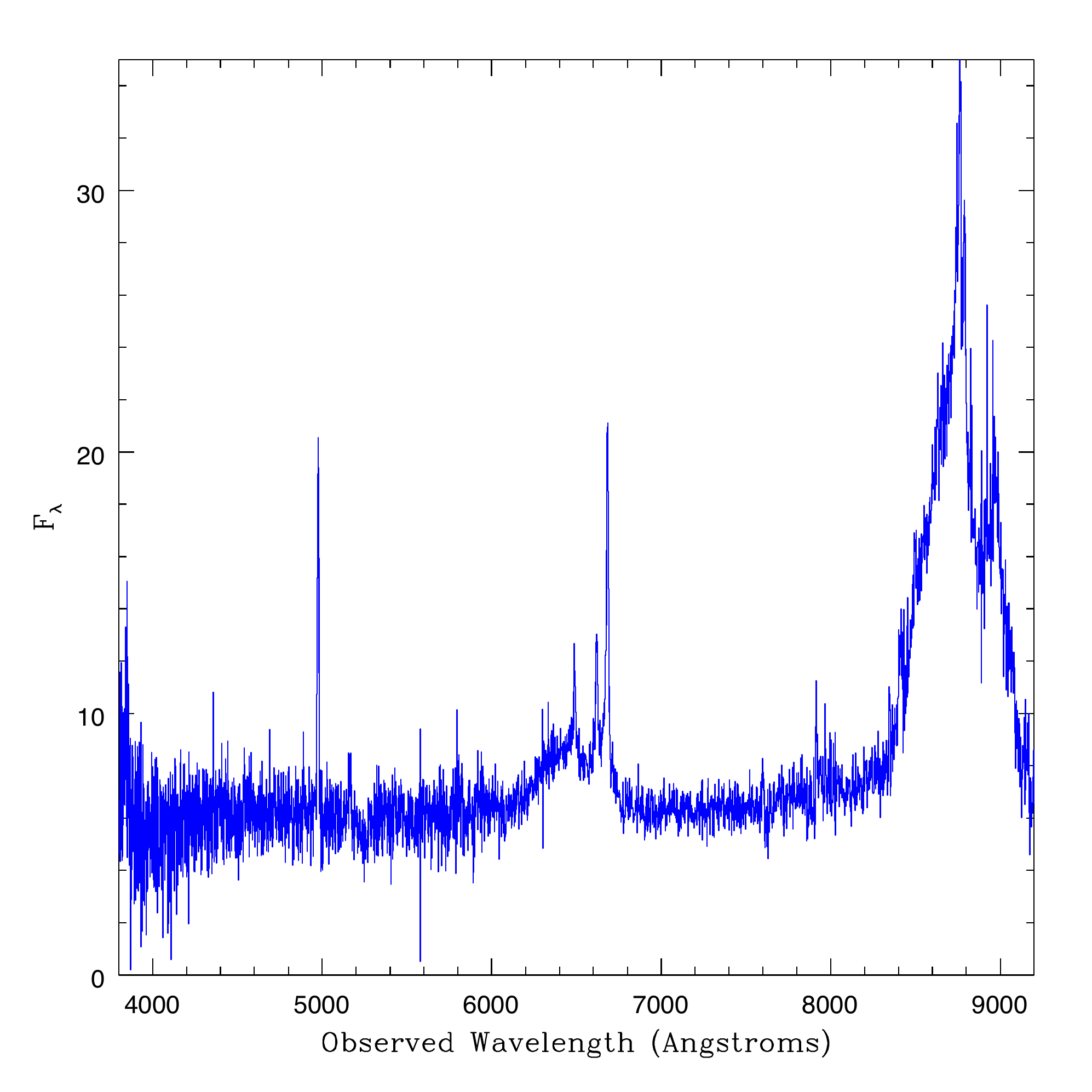}}
        \resizebox{9cm}{!}{\includegraphics{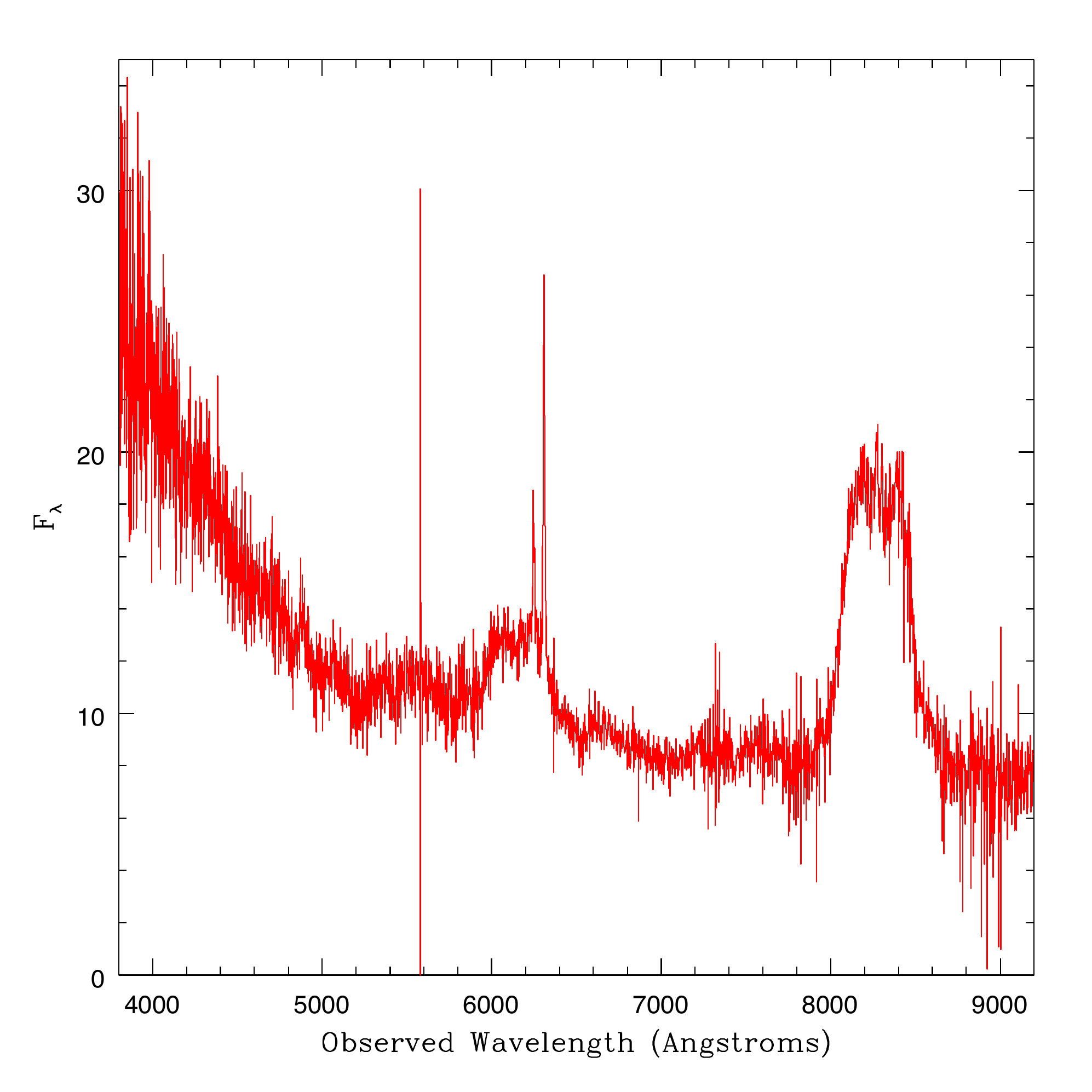}}\\
        \resizebox{9cm}{!}{\includegraphics{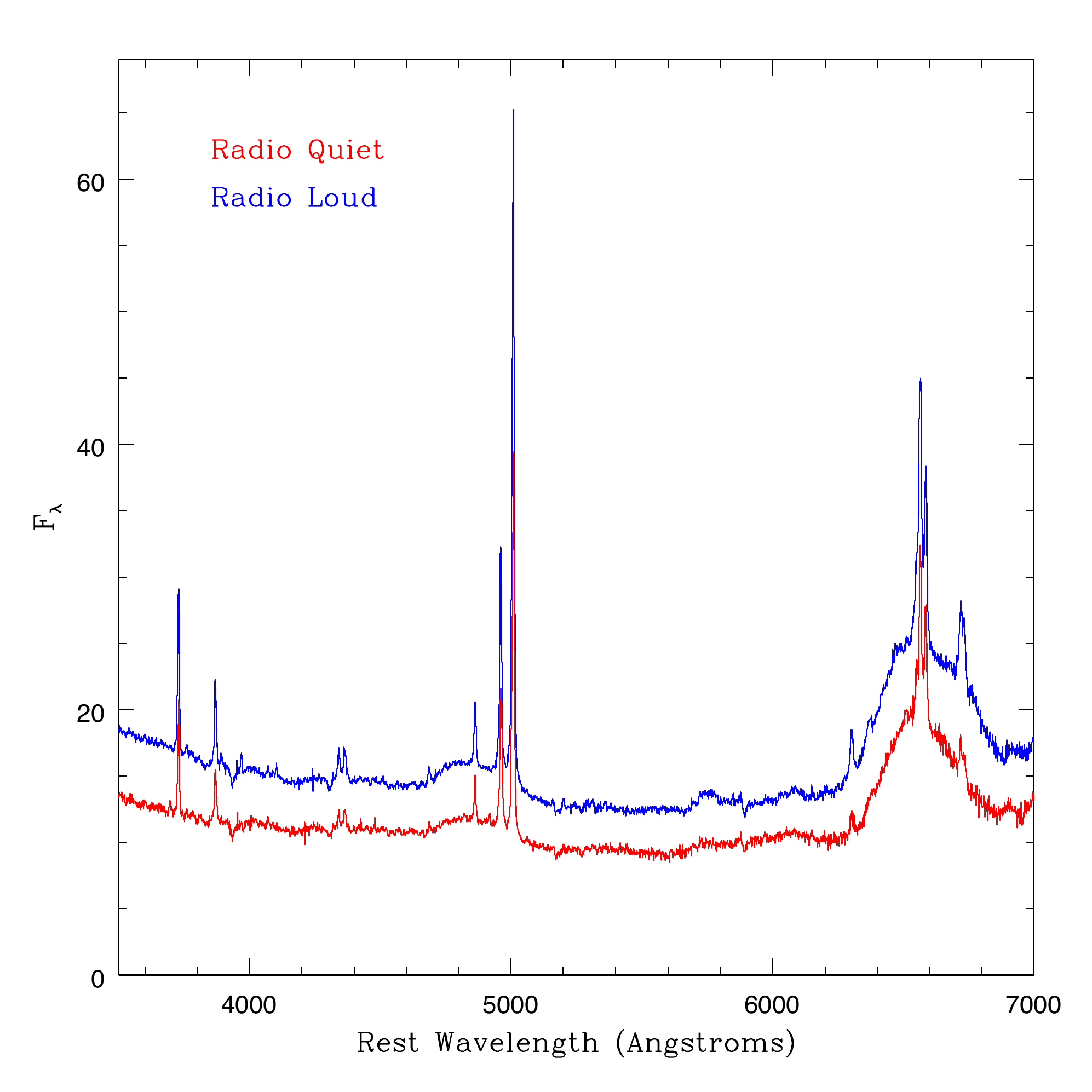}}
        \resizebox{9cm}{!}{\includegraphics{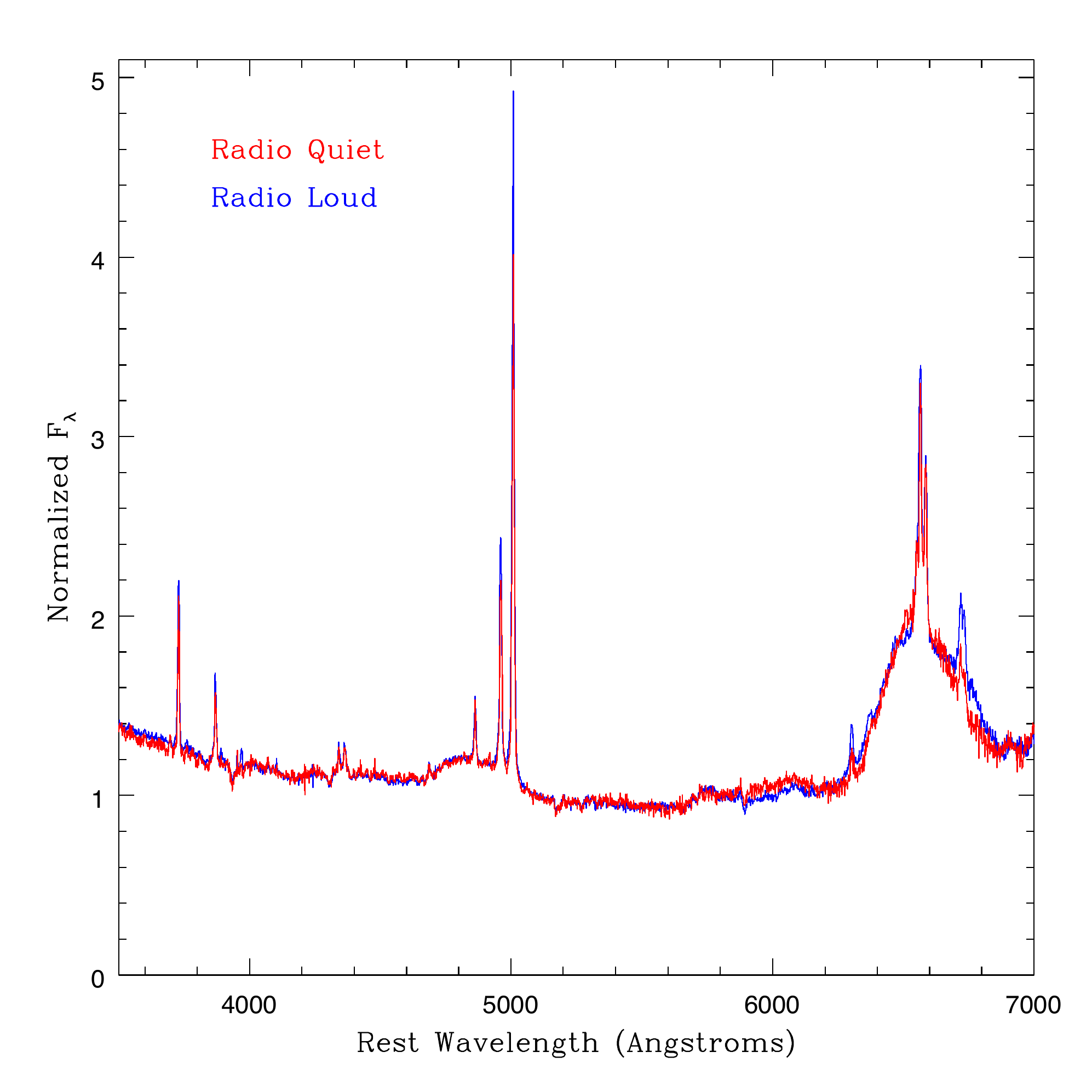}}\\
\end{tabular}
\caption{{\bf Top Left Panel:}Spectrum of a RL quasar with H$\beta$ FWHM greater than 15,000 km $s^{-1}$. $F_{\lambda}$ is in the units of $10^{-17}$ erg\,cm$^{-2}$\,s$^{-1}$\,\AA$^{-1}$. {\bf Top Right Panel:}Same as the top left panel but for a RQ quasar. {\bf Bottom Left Panel:}Combined spectra of RL and RQ quasars with H$\beta$ FWHM greater than 15000 km $s^{-1}$. $F_{\lambda}$ is in the units of $10^{-17} erg/cm^{2}/s/$\AA. {\bf Bottom Right Panel:} Normalised combined spectra of RL and RQ quasars with H$\beta$ FWHM greater than 15,000 km $s^{-1}$, where $F_{\lambda}$ is in the units of $10^{-16}$ erg\,cm$^{-2}$\,s$^{-1}$\,\AA$^{-1}$. 
From the combined spectra we can see that the RL quasars have higher optical continuum flux and stronger [O III] line. See section \S 3.2.} 
\label{fig:SDSSspectra}
\end{center}
\end{figure*}

\section{Results}

Here we present our results for the HBL as well as the entire H$\beta$ broad emission line sample of \citet{shen} for RL and RQ quasars.

\subsection{Radio Loud Fraction (RLF) and the HBL Sample}

 Figure \ref{fig:RLF} shows a plot of RLF \textit{versus} FWHM of broad H$\beta$ emission line of SDSS quasars. It clearly exhibits an increasing trend of RLF with the RLF reaching $\sim$57\% in the highest FWHM bin. We note that we see a similar trend when we use the MgII sample \citep{CBC21}. Additionally we found the correlation coefficient (Figure \ref{fig:RLF}) to be 0.97 with a $p$-value 0.0002, implying only a 0.02\% chance of finding such high value of the correlation coefficient by chance if the data were not correlated. While relative uncertainties in RLF are large in the three bins containing FWHM $> 10^4$ km\,s$^{-1}$ it is evident that the RLF value for the highest FWHM bin is 1-$\sigma$ above that containing FWHM value between $10^4$ and $1.25 \times 10^4$ km\,s$^{-1}$. That indicates the increase in RLF value at higher FWHM is significant. To investigate this phenomenon in more detail, particularly at the very high end of the H$\beta$ FWHM values, we analyse the properties of the HBL RL and RQ quasars. 
 
 In CB21, we showed that the distributions of the fiducial virial BH mass, optical continuum luminosity, and the bolometric luminosity of our HBL sample are slightly different from the parent catalog of \cite{shen}. 
 CB21 noted that the HBL RL quasars have higher optical continuum luminosity at 5100 \AA ~in the rest frame of the quasars as well as higher bolometric luminosities, compared to the RQ ones, while in the parent sample there is no difference between the luminosities and BH mass between RL and RQ quasars. There is a possibility that the optical continuum luminosity may be contaminated by light from the host galaxy. CB21 used a correction provided by \cite{shen}, who utilized an empirical fitting formula of the average host contamination by stacking spectra of quasars for that purpose.  
 
 In Figure \ref{fig:M-Ledd-CF} top panel, we show the distribution of BH mass for the RL and RQ quasars in our HBL sample. While computing the BH mass, unlike CB21, we include a correction due to $R_{Fe}$ \citep{du}. The $R_{Fe}$ corrected mass is a more robust and accurate estimate of mass compared to the ones obtained in DR7. The following equation defines the modified BH mass.
 \begin{equation*}
 {\rm M = f_{BLR} \frac{\Delta V^2 R_{BLR}}{G}}
 \end{equation*}
Where $f_{BLR}$ is the virial factor related to the geometry and kinematics of the BLR \citep[e.g.,][]{onken04, Ho14, Wo15} which is 0.4 for FWHM$\geq$ 10,000 km$s^{-1}$, $\Delta V$ is the FWHM and R$_{BLR}$ is the size of the BLR region, which is R$_{H\beta}$ in our case. The radius of the H$\beta$ emitting region in light-days can be written as  $$\rm {Log({R}_{H\beta}) = \alpha + \beta Logl_{44} + \gamma R_{Fe}},$$ Where $\alpha$ = 1.65 $\pm$ 0.06, $\beta$ =  0.45 $\pm$ 0.03, $\gamma$ = − 0.35 $\pm$ 0.08 and $\rm{l_{44} = L_{5100\AA}/10^{44}}$ erg/s \citep{du}.

Our results with the $R_{Fe}$ correction do not show any significant difference in the distribution of BH mass between the HBL RL and RQ quasar samples (we find a $p$-value for Wilcoxon rank sum test 0.0726, which implies that the null hypothesis that the two distributions are different may be rejected at the 95\% confidence level). We compute the Eddington ratios (Figure \ref{fig:M-Ledd-CF} middle panel) of our HBL sample, using our estimates of the $R_{Fe}$ corrected BH mass and bolometric luminosity. From the figure we see that there is almost no significant difference between mean values of the Eddington ratios of RL and RQ quasars ($p$-value 0.2317, which again implies that the null hypothesis that the two distributions are different may be rejected). 
 
 

\begin{figure}
\begin{center}
\begin{tabular}{c}
        \resizebox{9cm}{!}{\includegraphics{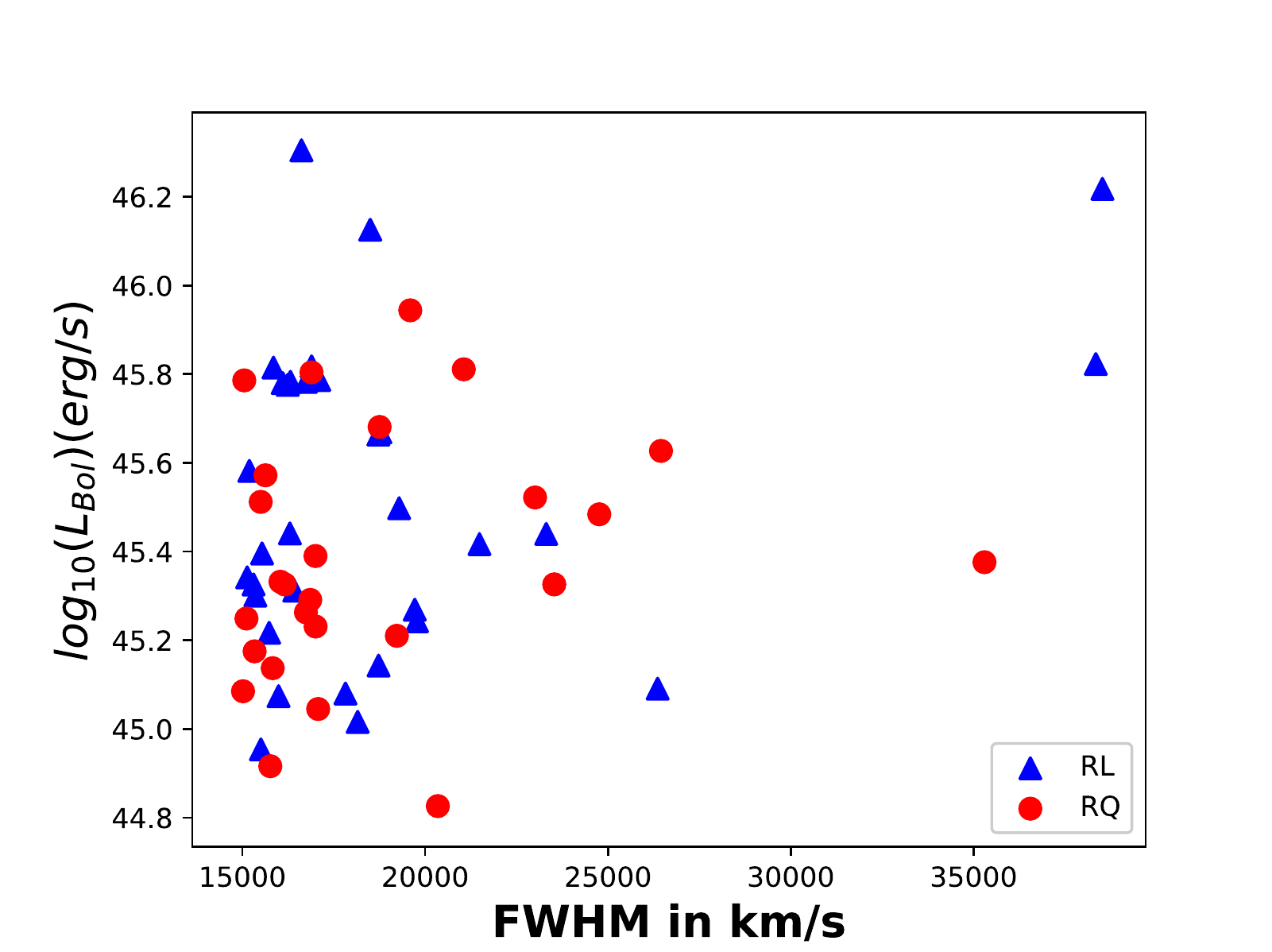}}\\
        \resizebox{9cm}{!}{\includegraphics{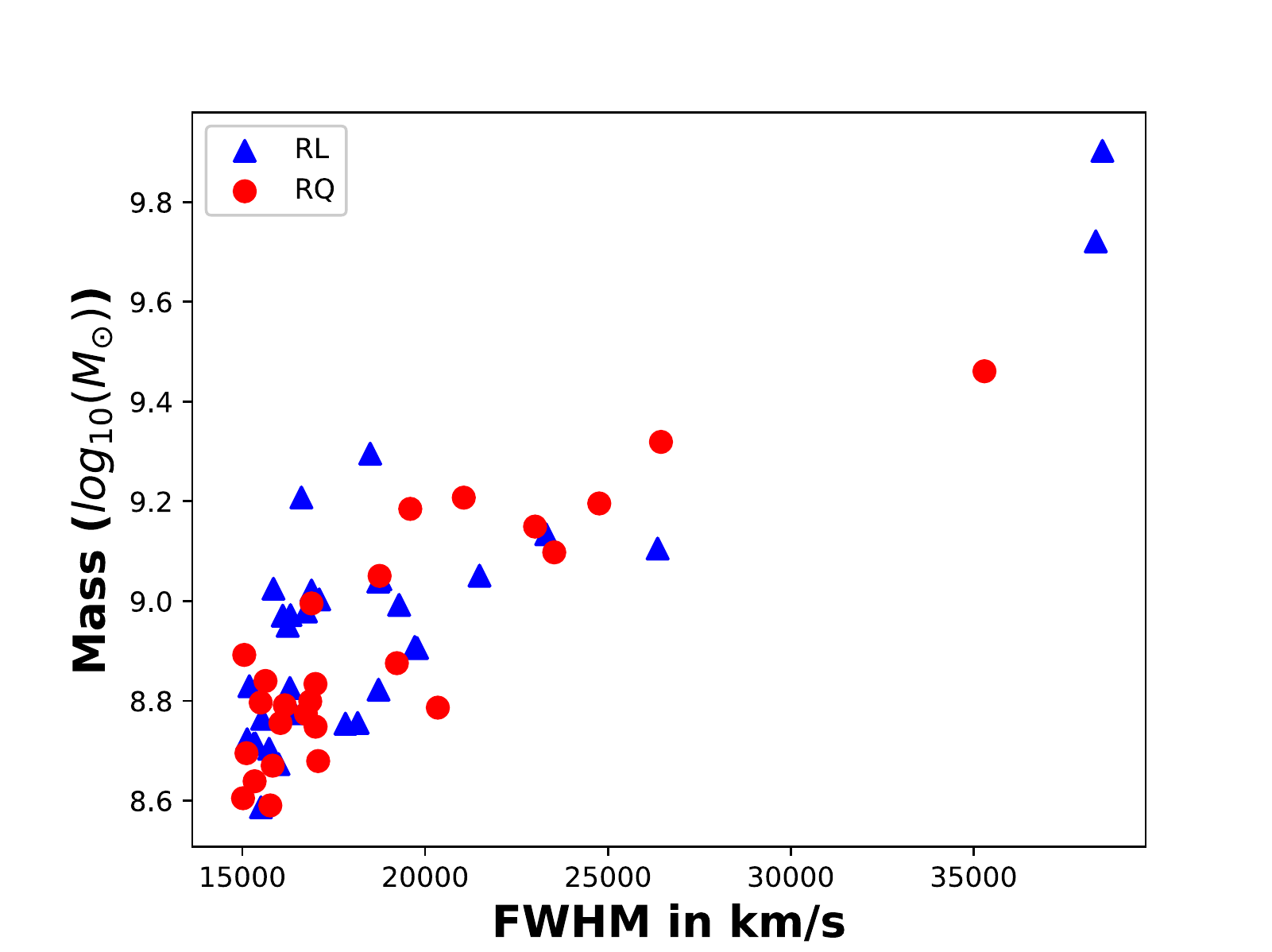}}\\
        \resizebox{9cm}{!}{\includegraphics{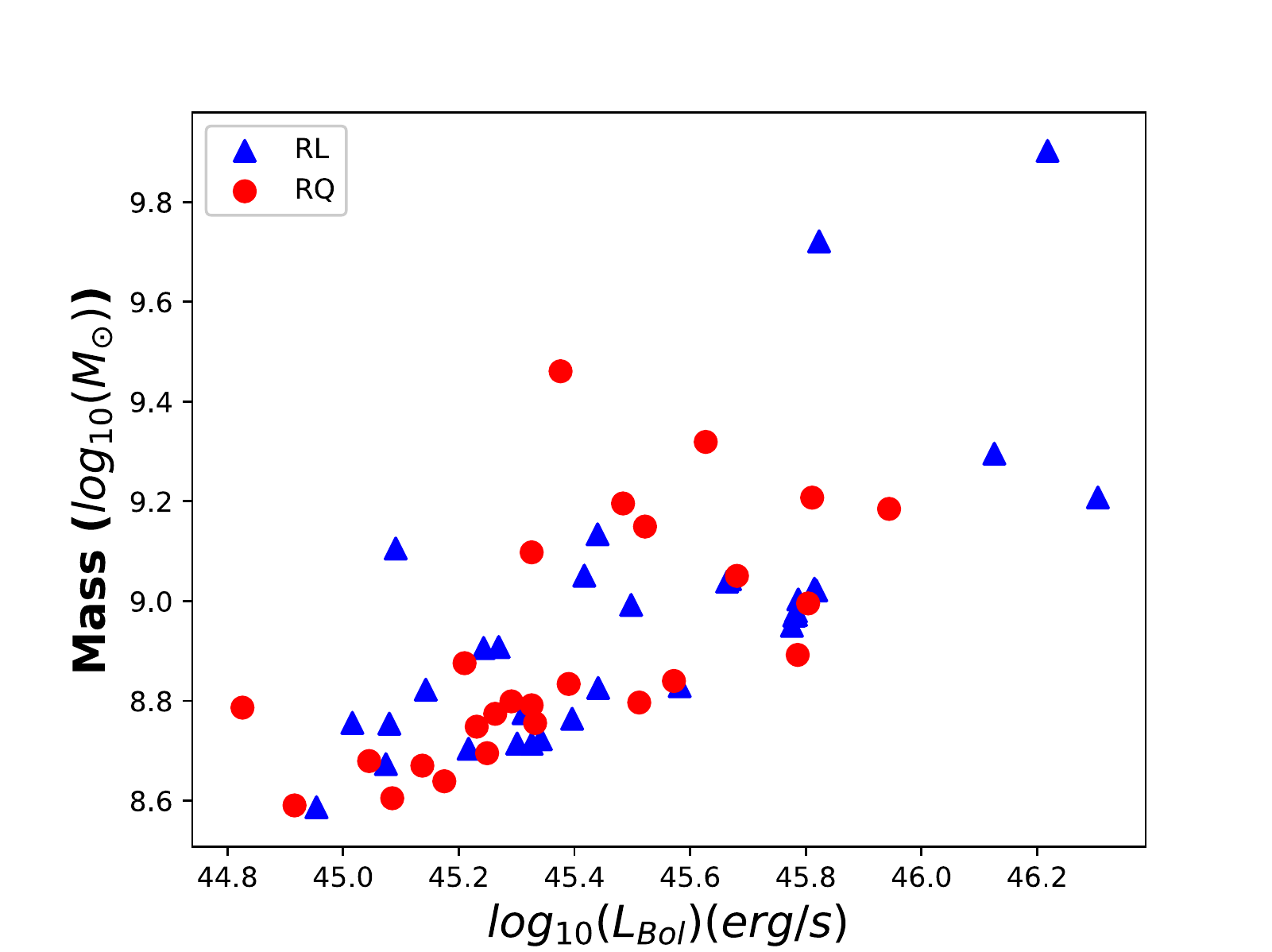}}\\
\end{tabular}
\caption{{\bf Top panel: }Red filled circles and blue filled triangles denote the RQ and RL sources in our HBL sample respectively on the bolometric luminosity \textit{versus} H$\beta$ FWHM plane. {\bf Middle panel: } Same as the top panel but shows the variation of the BH mass \textit{versus} H$\beta$ FWHM. We note that our black hole masses are computed using the $R_{Fe}$ correction. {\bf Bottom panel} Same as the other two panels but on the BH mass \textit{versus} bolometric luminosity plane. The related correlation matrices are shown in Tables 1 and 2}
\label{fig:correlation}
\end{center}
\end{figure}

It is known that, in a torus outside the BLR, the continuum and line emission are further modified by dust. UV and optical light are absorbed and scattered by dust and part of the energy is re-emitted at infrared wavelengths, causing anisotropic obscuration. 
This obscuration is often quantified by the covering fraction (CF).  We estimate the covering fraction from the ratio of the integrated flux at the mid-IR (MIR) to optical wavelengths, which is quantified as the fraction of the optical flux coming from the accretion disk, that gets reprocessed into the infrared by the dusty torus \citep[e.g.,][]{Netz}. We have taken $r-W1$ as the MIR flux, where $r$ is the SDSS $r$-band flux, corrected for galactic extinction and $W1$ is the $k$-corrected normalised (at $z=2$) flux at the WISE W1 band \citep{Yue}. The distributions of CF are shown in the bottom panel of Figure \ref{fig:M-Ledd-CF}. For the optical flux, $k-$corrected normalised (at $z=2$) optical continuum flux has been used in our analysis. 

For our HBL sample we see that there is no difference in the mean values of CF distribution of RL and RQ quasars ($p$-value 0.6890 with usual significance). In the local universe the number ratios of obscured to unobscured active galactic nuclei (AGNs) provide useful constraints on the average covering fraction of the torus \citep{reyes08, lu10, ma&wang13}. If, for instance, RL quasars have smaller CF than RQ, we will be observing them at larger angles. In that case the FWHM of RL quasars may be higher due to the differences in angular distributions. Our analysis of the CF reveals that for our HBL sample, orientation is not a critical factor. 

\subsection{Comparison of Properties of the RL and RQ quasars}

We have shown the dependencies of the different properties of RL quasars in the HBL sample in the left and right panels of Figure \ref{fig:Lrad-L5100-R-M}. The variables we use are, $L_{\rm rad}$, $L_{5100\AA}$, BH mass and $R$, which imply radio luminosity at 6 cm, optical continuum luminosity at $5100$ \AA, black hole mass and radio loudness (ratio of radio luminosity to optical continuum luminosity), respectively. Errors associated with $L_{\rm rad}$ and $R$ are $\sim$ 1\% and BH mass are $\sim$ 35\% to 42\%. Optical luminosity errors are quoted from \cite{shen} and BH mass, $L_{\rm rad}$ and $R$ errors are calculated accordingly. We note that RQ quasars may generate weak but non-negligible radio emission through nuclear as well as star formation processes. \citep[e.g.,][]{kim,heck&best14,zam16,sym17}. They may contain low-power jets although it may not contribute significantly to the overall energy budget \citep[e.g.,][]{harr14,jar19}. However, we follow \cite{rich}, who use the FIRST completeness limit. To maintain the limiting radio luminosity, only the RL quasars have been considered in our analysis.

For our HBL sample we see a correlation between $L_{rad}$ and L$_{5100\AA}$ such that a linear regression relation may be obtained given by $L_{rad} = (1.36 \pm 0.30)$ L$_{5100 \AA} -(28.89 \pm 13.58)$ with a correlation coefficient of 0.62. However, we note that if a pair of luminosities are correlated that could be driven by their dependency on redshift. To test the reliability of the correlations obtained, we calculate partial correlation coefficient \citep{akri&sie95} to check the dependence of the radio luminosity and optical continuum luminosity at $5100$ \AA ~ on redshift. Pearson’s partial correlation coefficient (PPCC) eliminates the common dependence on a third variable and measures the correlation between two quantities. We obtain PPCC $= 0.26$ for radio and optical continuum luminosity, which is significantly lower than the correlation coefficient between them implying the correlation to be much weaker when the redshift dependency is subtracted.
On the other hand, we do not find any significant dependence of $R$ on the BH mass. We have verified that for a given continuum luminosity the radio luminosity is higher for the HBL quasars compared to the non-HBL ones. 

We compute the composite spectra of RL and RQ quasars of our HBL sample. We first obtain the individual spectra from SDSS (Figure \ref{fig:SDSSspectra}, upper panels), de-redshifted them and combine the rest frame RL and RQ spectra separately. We compare them on the bottom left panel of Figure \ref{fig:SDSSspectra}. We find that the average continuum flux is higher for RL quasars.
From the normalised composite spectra (Figure \ref{fig:SDSSspectra}, bottom right) we observe that one of the major differences between the combined spectra of RL and RQ quasars is in the strength of the [O III] narrow line (equivalent width for RL is $42.9\pm0.1$ and for RQ is $41.7\pm0.2$) while the H$\beta$ broad lines are identical for both the samples. So the difference 1.2 and error associated with the difference is 22\% (less than 1-$\sigma$). We checked that these effects are physical effects and not driven by the most massive sources.

\begin{table}
\caption{Correlation Matrix for RL quasars of our HBL Sample} 
\centering 
\begin{tabular}{c c c c} 
\hline\hline 
   & $L_{Bol}$ & BH mass & FWHM \\ [0.5ex] 
\hline 
$L_{Bol}$ & 1.0000 & 0.6855 & 0.0829\\
BH mass & 0.6855 & 1.0000 & 0.7757\\ 
FWHM & 0.0829 & 0.7757 & 1.0000\\ [1ex]
\hline 
\end{tabular}
\end{table} 

\begin{table} 
\caption{Correlation Matrix for RQ quasars of our HBL Sample} 
\centering 
\begin{tabular}{c c c c} 
\hline\hline 
 & $L_{Bol}$ & BH mass & FWHM\\ [0.5ex] 
\hline 
$L_{Bol}$ & 1.0000 & 0.5352 & -0.1464\\
BH mass & 0.5352 & 1.0000 & 0.7486\\ 
FWHM & -0.1464 & 0.7486 & 1.0000\\ [1ex]
\hline 
\end{tabular}
\end{table}

\subsection{Correlation of Various Properties of the HBL Quasars} 
 We perform correlation studies between bolometric luminosity, FWHM, and BH mass of our HBL sample where the errors associated with bolometric luminosity and FWHM are $\sim$ 1\% and $\sim$ 20\% respectively \citep{shen}. The results are summarized in Tables 1 and 2. 
 In our sample we observe, as shown in Figure \ref{fig:correlation}, that bolometric luminosity is not correlated with FWHM ($p$-value for RL and RQ are 0.15 and 0.39 respectively) but BH mass is correlated with FWHM ($p$-value 0.0001 for both RL and RQ) as well as bolometric luminosity ($p$-value for RL and RQ are 0.0001 and 0.0004). Both FWHM and bolometric luminosity increase with BH mass for both our RL and RQ quasar samples. We have found that the four most massive and the most luminous sources (upper right corner : Figure \ref{fig:correlation}, bottom panel) are RL quasars. As evident from the tables, the correlation between bolometric luminosity and BH mass is slightly stronger in RL quasars compared to the RQ ones. 
 
\begin{figure}
\begin{center}
\begin{tabular}{c}
        \resizebox{8cm}{!}{\includegraphics{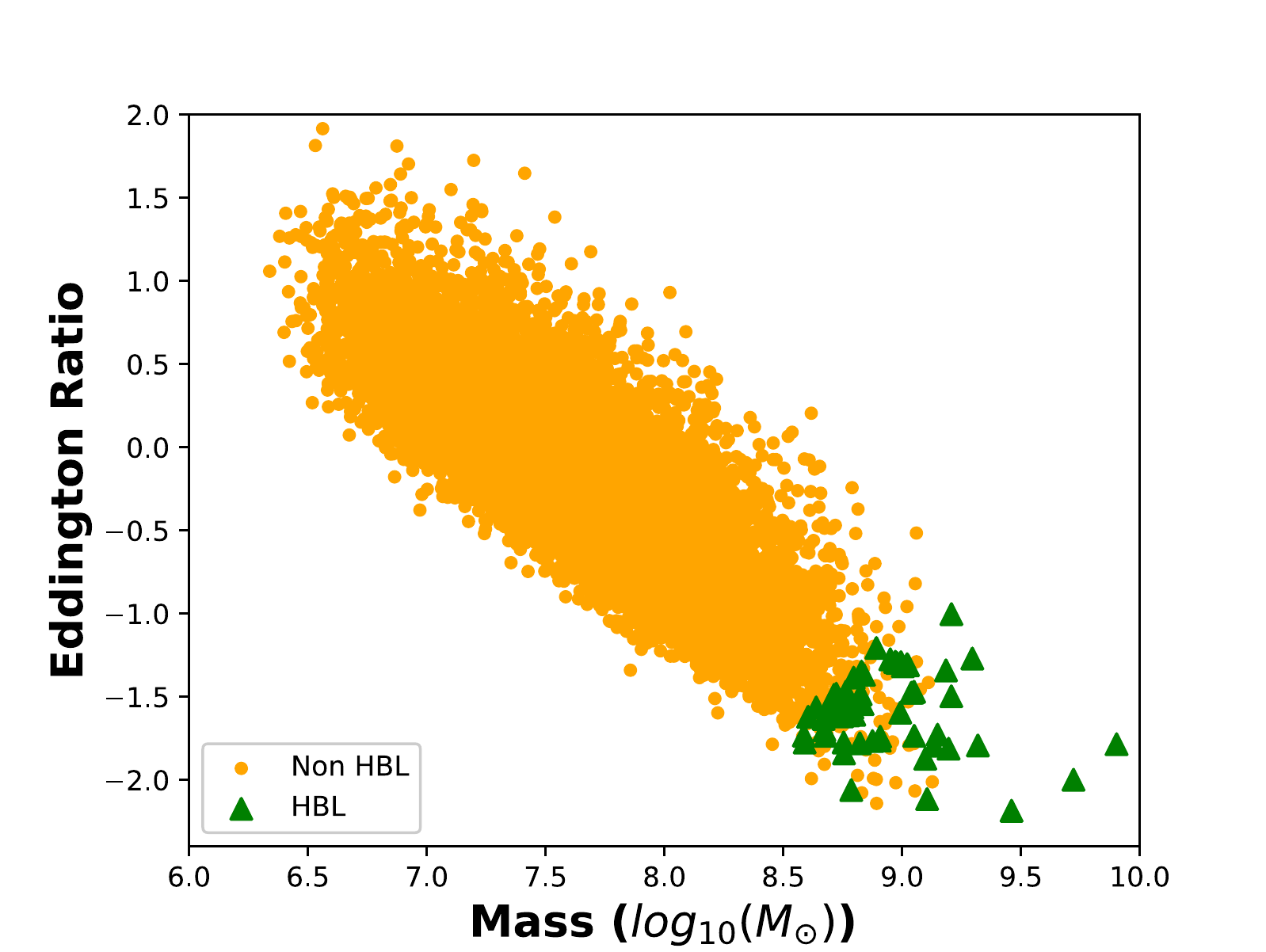}}\\ 
        \resizebox{8cm}{!}{\includegraphics{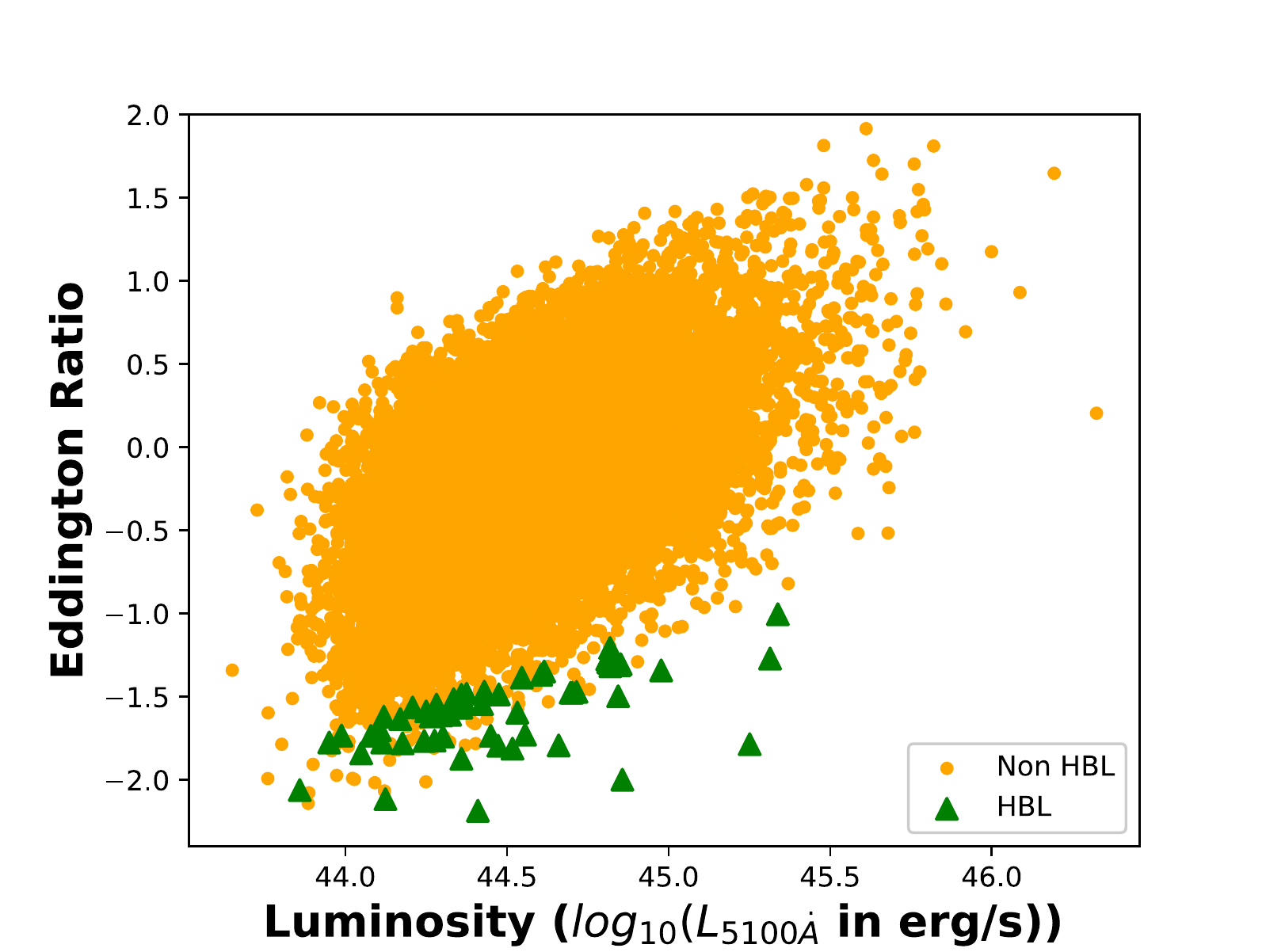}}\\ 
        \resizebox{8cm}{!}{\includegraphics{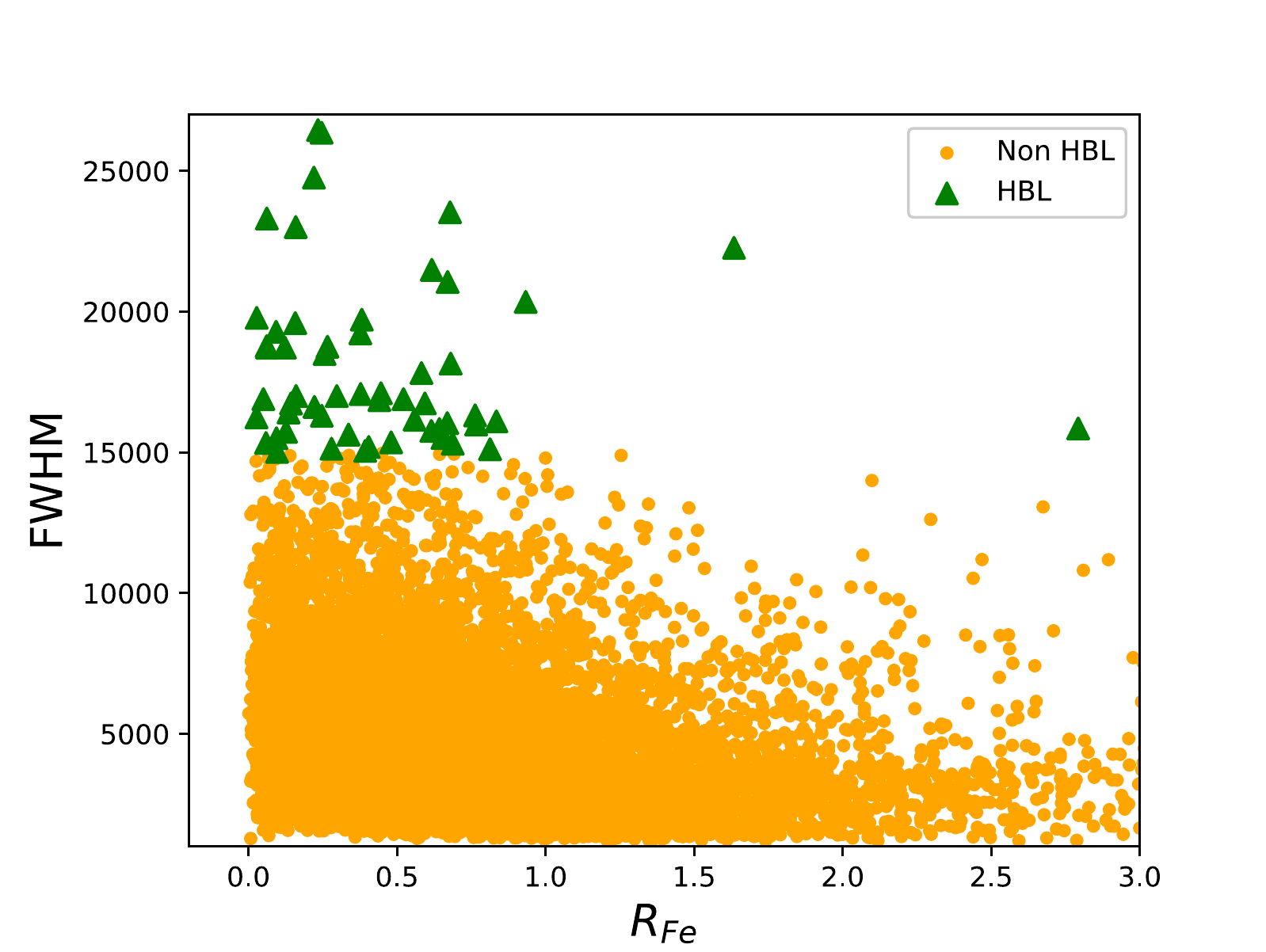}}\\
\end{tabular}
\caption{{\bf Top Panel: }Yellow filled circles and green filled triangles show the variation of Eddington ratio with BH mass of quasars with H$\beta$ FWHM greater and less than 15,000 km $s^{-1}$, i.e., non-HBL and HBL, respectively. {\bf Middle Panel: }Same as the top panel but here the horizontal axis denotes the optical continuum luminosity. {\bf Bottom Panel: } Same as the above two panels but shows the variation of H$\beta$ FWHM \textit{versus} $R_{Fe}$. As we know that the Eddington ratio is inversely related to the BH mass, we note that our HBL sample contains the quasars with the highest BH mass and the lowest Eddington ratios. In addition, it demonstrates that the bolometric luminosity is not correlated with the Eddington ratio for the total sample. Furthermore, we can see $R_{Fe}$ is inversely related to FWHM for the total sample and the quasars in our HBL sample has the lowest $R_{Fe}$ values. See \S 3 and 4 for more discussions.}
\label{fig:non-HBL}
\end{center}
\end{figure}

\subsection{Comparison of the HBL and Non-HBL Sample} 
We compare our HBL sample with quasars from the non-HBL sample (Figure \ref{fig:non-HBL}). The top, middle, and bottom panels show the distributions of mass versus Eddington ratio, Eddington ratio versus optical continuum luminosity, and the FWHM versus $R_{Fe}$, relations respectively, of both the samples of interest. Errors associated with Eddington ratio and $R_{Fe}$ are $\sim$ 40\% and $\sim$ 1.5\% respectively. We find that quasars in our HBL sample has the highest BH mass and lowest Eddington ratios and optical continuum luminosities compared to the non-HBL sample. We also observe that our HBL sample has the lowest $R_{Fe}$. For the entire sample we found $R_{Fe}$ to be decreasing with FWHM consistent with previous studies \citep{zfir, Yue}. 

\section{Summary and Discussion}

In Figure \ref{fig:RLF}, we show that the radio loud fraction increases with the FWHM of broad H$\beta$ emission line for z $< 0.75$ quasars identified in the SDSS. To further investigate this effect, we selected a subsample of extreme objects (FWHM $>$ 15,000 km $s^{-1}$) to study their properties in detail. For our sample, number of RL quasars is 56 among a total of 97, which implies an RLF of 57\%, much larger than that of the larger sample of all SDSS quasars matched with FIRST.On the other hand, the BH mass of the quasars in our subsample is, on average, higher than that of the non-HBL SDSS quasars while the optical continuum luminosities are lower on average. Consequently, the Eddington ratio of the quasars in our subsample  $\sim 10^{-2} - 10^{-1}$, is significantly lower than the average of that of the non-HBL quasars. Thus, for equivalent BH mass range, quasars in our sample tend to have lower luminosity and hence a lower accretion rate.

It has been observed in stellar mass BH accreting systems that the X-ray spectrum becomes harder when the luminosity goes below a threshold, the value of which is in the range $0.01-0.1$ $L_{edd}$, where $L_{edd}$ is the Eddington luminosity. It has been thoroughly discussed in the literature that this hard state may be caused by a switch of the nature of the accretion flow from the standard geometrically thin optically thick radiatively efficient disk to a geometrically thick optically thin advection-dominated accretion flow (ADAF) \citep{nar1, nar2, abr, kato, nar3}. In the ADAF, unlike a standard disk, heat produced in the accretion through viscous dissipation is not emitted. Instead the energy is confined to the accreted material and moves toward the BH along with the flow. As a result the accretion luminosity is low \citep{nar2, naka, nar3}. ADAF has been used to explain observations of various classes of AGN as well. ADAF has the additional property that the accreted material is not strongly bound to the BH. Hence, there is a higher probability of producing disk wind and jet in AGN containing an ADAF \citep{mei, nar3}. Although our findings show some characteristics of ADAF, we note that ADAF models are unlikely to produce the strong optical emission observed in quasars. Studies of jet systems led to magnetically arrested accretion \citep[e.g.,][]{tche11}. We refer the reader to \citet{hard20} for discussion on accretion processes in accreting systems with radio jets. We propose that our subsample of quasars with high FWHM of emission lines may contain magnetically arrested accrection flows, which is consistent with having a very large fraction being radio-loud.

In our subsample of high broad line quasars, there is no significant difference in mass among the two classes, namely, radio-loud and radio-quiet. However, the RL quasars have larger bolometric luminosity than the RQ ones as computed by \cite{shen}. Furthermore, we computed the combined spectrum of SDSS quasars with FWHM $>$ 15,000 km $s^{-1}$. It shows the mean optical continuum luminosity to be higher for RL quasars, in comparison to RQ quasars. Given that the difference in the bolometric or optical continuum luminosity is mainly due to that of the accretion disk, the above implies some connection between a stronger jet and a brighter disk. Accretion disc-jet connection in AGN has often been studied using X-ray and radio observations \citep[e.g.,][]{naik, marscher2002, RC1, chatterjee2011, RC2}. In AGN, X-rays are not produced in the disk itself but in a distribution of energetic particles located near the disk termed corona, which is assumed to be energetically and dynamically connected to the disk. On the other hand, optical emission is directly produced in the disk and hence may be better suitable for probing its connection to the jet. \cite{Ino} have found a strong correlation between the disc and the jet, using bolometric luminosity (as inferred from the optical luminosity of the disc) and 1.4 GHz radio luminosity as an indicator of the jet emission. We note that our findings are consistent with this result although our sample size is not very large.

We found that the emission line properties of the composite spectra of the RL and RQ quasars of our HBL sample are similar. The major difference between the two classes has been observed in the strength of the [O III] narrow line. The composite spectrum of the HBL RL quasars exhibits a stronger [O III] line compared to that of the RQ ones. This is consistent with previous findings that RL quasars have a larger narrow line region and stronger narrow emission lines than their RQ counterparts \citep[e.g.,][]{heckman1991,brotherton}. 
This may be caused by the interaction of the radio jet with the narrow line gas, as inferred from observations indicating an alignment between the radio jet axis and the narrow emission line region \citep[e.g.,][]{cimatti1993}.  On the other hand, \cite{schul} have shown an enhancement in [O III] line strength by a factor of at least 1.5 in a radio-loud compared to a radio-quiet sample, both of which have similar redshifts, black hole mass, and accretion rate as indicated by the optical continuum luminosity. They conclude that the enhanced [O III] line strength is due to higher extreme UV luminosity at 35 eV for the RL objects. They suggest this is caused by the radio-loud quasar population having systematically higher black hole spin and consequently their radiative efficiency being larger, consistent with the paradigm \citep[][]{wilson1995,sikora,tchek10} that the radio loud objects have higher black hole spin. 

In this work we employ a sample of high broad line quasars and search for hints in the observed higher radio loud fraction in these objects compared to the general population of quasars. Our study suggests that a magentically arrested flow like accretion and a stronger connection between the disc and the jet may result in this higher radio function. We propose that this might provide a clue to the broader question of understanding radio dichotomy in the quasar population.

\section*{Acknowledgement}
AC thanks Jaya Maithil from the University of Wyoming,  Swamtrupta Panda from the Laborat{\'o}rio Nacional de Astrof{\'i}sica - MCTIC and Preeti Kharb of National Centre for Radio Astrophysics, for their valuable inputs. The authors thank the referee for several important feedback which greatly helped in improving the draft. 

SC acknowledges support from the Department of Science and Technology through the SERB-ECR grant, SERB-CRG grant and from the Department of Atomic Energy for the BRNS grant. RC and SC are grateful to the Inter University Center for Astronomy and Astrophysics (IUCAA) for providing infrastructural and financial support along with local hospitality through the IUCAA-Associateship program. RC thanks Presidency University for support under the Faculty Research and Professional Development (FRPDF) Grant, and ISRO for support under the AstroSat archival data utilization program. RC acknowledges financial support from BRNS through a project grant (sanction no: 57/14/10/2019-BRNS) and thanks the project coordinator Pratik Majumdar for support regarding the BRNS project.

\section*{Data availability}
The data underlying this article will be shared on reasonable request to the corresponding author.

\bibliographystyle{mnras}
\bibliography{arxiv}

\end{document}